\documentclass[twocolumn,showpacs,prc,amsmath,amssymb,nofootinbib]{revtex4}

\usepackage{color}
\usepackage{graphicx}
\usepackage{dcolumn}
\usepackage{bm}
\usepackage{lscape}
\usepackage{mathrsfs}
\usepackage{braket}
\usepackage{txfonts}

\begin{document}


\title{Investigation of 3/2$_2^-$ state of $^{9}$Li nucleus\\
with microscopic structure and reaction models}

\author{T.~Furumoto}
\email{furumoto@ichinoseki.ac.jp}
\affiliation{Ichinoseki National College of Technology, Ichinoseki, Iwate 021-8511, Japan}

\author{T.~Suhara}%
\email{suhara@matsue-ct.ac.jp}
\affiliation{Matsue College of Technology, Matsue, Shimane 690-8518, Japan}

\author{N.~Itagaki}%
\email{itagaki@yukawa.kyoto-u.ac.jp}
\affiliation{Yukawa Institute for Theoretical Physics, Kyoto University, Kyoto 606-8502, Japan}

\date{\today}

\begin{abstract}
The low-lying states of the $^{9}$Li nucleus are investigated with a unified framework of microscopic structure and reaction models.
In the structure model, the wave function is fully antisymmetrized and the $^{9}$Li nucleus is described as an $\alpha$ + $t$ + $n$ + $n$ four-body system, and low-lying 1/2$^{-}$, 3/2$^{-}$, 5/2$^{-}$, and 7/2$^{-}$ states are obtained by the stochastic multi-configuration mixing method.
Using these wave functions, the quasi-elastic cross section at $E/A$ = 60 MeV and the elastic and inelastic cross sections at $E/A$ = 50 MeV on the $^{12}$C target are calculated in the framework of the microscopic coupled channel (MCC) method.
The characteristic inelastic angular distribution is seen in the 3/2$_{2}^{-}$ state, whose $\alpha$+$t$ cluster structure and valence neutron configurations are discussed in detail.
We find the possibility of triaxial deformation and mixing of di-neutron components in the $^{9}$Li nucleus.
\end{abstract}


\maketitle

\section{Introduction}
The study of the unstable nuclei has been extensively carried out, since the light unstable nuclei are considered to play a key role in determining the abundance ratio of elements heavier than Fe.
Much efforts have been devoted because of the interests not only to the nucleosynthesis  but also to the structure of nuclei itself.
The unstable nuclei have been known to have exotic properties, which have never been seen in stable nuclei.
For instance, the discovery of the halo structure and the change of the magic number(s) have been reported~\cite{TAN85, OZA00}.
The halo nucleus consists of a core nucleus and valence nucleon(s), and the valence nucleon(s) is bound with tiny binding energy, which makes the spatial extension of the valence nucleon(s) very large.
As a result, the halo nucleus has a large radius, which can be observed as a large total reaction cross section in the reaction experiments~\cite{TAN85}.

For the study of the halo nucleus, the importance of the core excitation effect has been recently reported in the $^{8}$B elastic scattering on the $^{12}$C target~\cite{HOR10}. 
In addition, Moro and his collaborators have shown such important role of the core excitation effect of the $^{11}$Be nucleus in the inelastic and breakup cross sections~\cite{CRE11, MOR12, MOR12-2}. 
Both of the $^{8}$B and $^{11}$Be nuclei have ``a valence nucleon + core nucleus" structure ($p$ + $^{7}$Be and $n$ + $^{10}$Be structure for $^{8}$B and $^{11}$Be, respectively).
Here the separation energy of the valence nucleon is only the order of several hundred keV.
To understand the reaction mechanism of such weakly bound systems, we have to focus on the effect of not only the breakup of the system into a valence nucleon and a core nucleus but also the core excitation. 
Especially in the elastic scattering cross section, it has been studied that the core excitation effect, rather than the breakup effect, plays a dominant role~\cite{HOR10}. 
It is indispensable to investigate the properties of the core nucleus and take the excited states into account in the reaction calculation. 
Here, the $^{11}$Li nucleus is well known to have two-neutron halo structure.
The excitation effect of the core part of the $^{11}$Li nucleus, $^{9}$Li, may play an important role to describe the nuclear reactions; however $^{9}$Li is also an unstable nucleus and its properties are not necessarily known well. 
Therefore, in this paper we aim to investigate the properties of low-lying states of the $^{9}$Li nucleus itself from the view point of the both of the nuclear structure and reaction.

So far there have been many challenges for the description of the structures of the $^{9}$Li nucleus. 
From the shell model side, the so-called {{\it ab initio}} approach, such as the no-core shell model calculation, is feasible for the low-lying states~\cite{NAV98}. 
Also the contribution of the tensor force is investigated with the tensor optimized shell model calculation~\cite{MYO12}. 
In addition, the $^{9}$Li nucleus is described with the $\alpha + t + n + n$ cluster model~\cite{ARA01}, which also well reproduces the properties of the low-lying states in the comparison with the experimental data. 
The exotic excited state with three triton clusters has been predicted by the stochastic multi-configuration mixing method with the $\alpha + t + n + n$ and $t + t + t$ configurations~\cite{MUT11}. 
The $^{6}$He + $t$ cluster structure has been also predicted with the Generator Coordinate Method (GCM) calculation~\cite{ENY12}. 

In this paper, the low-lying states of the $^{9}$Li nucleus is investigated using a unified framework of microscopic structure and reaction models.
Here, the ``unified" framework means that we investigate the structure and reaction of the $^{9}$Li nucleus with the same wave function, though the effective nucleon-nucleon ($NN$) interactions used in the structure and reaction parts are different.
The structure of the $^{9}$Li nucleus is described with the microscopic $\alpha + t + n + n$ cluster model, where the wave function is fully antisymmetrized.
The total wave function is a superposition of Brink-type basis states, and the configuration of each basis state, the positions of two clusters and two valence neutrons, is randomly generated based on the so-called stochastic multi-configuration mixing method~\cite{ICH11, MUT11}.
This cluster components are analogous to the $\alpha + \alpha + n + n$ model for $^{10}$Be, though $t$ cluster has a spin 1/2 in the $^{9}$Li case. 
Using the wave function, the inelastic cross sections are calculated to investigate the excited and resonance states of the $^{9}$Li nucleus in the framework of the microscopic coupled channel (MCC) method with complex $G$-matrix interaction CEG07~\cite{CEG07, FUR09}.
In this study, special attention is paid to the calculated inelastic cross sections of the 3/2$^{-}$ states, where the competition between the monopole and quadrupole transitions in the excitation from the ground $3/2^-$ state is discussed.
We also investigate the multi-step coupling effect on the inelastic cross section for the 3/2$^{-}_{2}$ state.
Finally, we discuss the possibility of triaxial deformation and mixing of di-neutron components in the $^{9}$Li nucleus.

\vspace{3mm}
\section{Formalism}
\subsection{Microscopic cluster model}
First, we start with the nuclear structure calculation part.
We introduce basis states with various $\alpha + t + n + n$ configurations, $\{ \Psi_{i}^{J^{\pi} MK} \}$, to describe the $^{9}$Li structure in the microscopic cluster model.
It is known that the ground state of the $^{9}$Li nucleus is well described by the $\alpha + t + n + n$ model~\cite{ARA01}, thus we apply this cluster model space to the low-lying excited states of $^{9}$Li in the first order approximation. 
The total wave function $\Phi^{J^{\pi} M}$ is therefore,
\begin{equation}
\Phi^{J^{\pi} M} = \sum_{K} \sum_{i} c_{i, K} \Psi_{i}^{J^{\pi} MK}.
\label{mcmwf}
\end{equation}
The eigenstates of Hamiltonian are obtained by diagonalizing the Hamiltonian matrix, and the coefficients $\{c_{i, K} \}$ for the linear combination of Slater determinants are obtained.

The $i$-th basis state of $\{ \Psi_{i}^{J^{\pi} MK} \}$ with the $\alpha + t + n + n$ configuration has the following form,
\begin{eqnarray}
\Psi_{i}^{J^{\pi} MK} = &&  P^{\pi} P^{JMK} {\cal A} \nonumber \\
&& \big[ \phi_{\alpha} (\bm{r}_{1} \bm{r}_{2} \bm{r}_{3} \bm{r}_{4}, \bm{R}_{1}) \phi_{t} (\bm{r}_{5} \bm{r}_{6} \bm{r}_{7}, \bm{R}_{2}) \nonumber \\
&& \ \phi_{n^{(1)}} (\bm{r}_{8}, \bm{R}_{3}) \phi_{n^{(2)}} (\bm{r}_{9}, \bm{R}_{4}) \big]_{i},
\end{eqnarray}
where $\cal A$ is the antisymmetrizer, and 
$\phi_{\alpha} (\bm{r}_{1} \bm{r}_{2} \bm{r}_{3} \bm{r}_{4}, \bm{R}_{1})$, 
$\phi_{t} (\bm{r}_{5} \bm{r}_{6} \bm{r}_{7}, \bm{R}_{2})$,
$\phi_{n^{(1)}} (\bm{r}_{8}, \bm{R}_{3})$,  
$\phi_{n^{(2)}} (\bm{r}_{9}, \bm{R}_{4})$ are wave functions of $\alpha$, triton, the first valence neutron, and the second valence neutron, respectively.
Here, $\{ \bm{r}_{i} \}$ represents spatial coordinates of nucleons, and each nucleon is described as locally shifted Gaussian centered at $\bm{R}$ ($\exp[-\nu(\bm{r}_i - \bm{R})^2]$) with the size parameter of $\nu = 1/2b^2$, $b=$ 1.46 fm.
The $\alpha$ cluster consists of four nucleons (spin-up proton, spin-down proton, spin-up neutron, and spin-down neutron), which share a common Gaussian center parameter $\bm{R}_{1}$, though the spin and isospin of each nucleon are not explicitly described in this formula for simplicity. 
The triton consists of three nucleons (proton, spin-up neutron, and spin-down neutron), which are centered at $\bm{R}_{2}$.
The Gaussian center parameters of two valence neutrons are $\bm{R}_{3}$ and $\bm{R}_{4}$.
The $z$ components of the spins of the two valence neutrons are introduced to be parallel or anti-parallel dependent on the basis state.
The index $i$ in Eq. (2) specifies a set of Gaussian center parameters for $\bm{R}_{1}$, $\bm{R}_{2}$, $\bm{R}_{3}$, and $\bm{R}_{4}$, and spin directions of valence neutrons.
The projection onto an eigenstate of parity and angular momentum operators (projection operators $P^{\pi}$ and $P^{JMK}$) is performed numerically. 
The number of mesh points for the Euler angle integral is  $16^3 = 4096$.
The value of $M$ specifies the $z$ component of the angular momentum in the laboratory frame, and the energy does not depend on $M$; however, the energy depends on $K$, which is a $z$ component of the angular momentum in the body-fixed frame.

The Hamiltonian operator $(H)$ has the following form:
\begin{equation}
\displaystyle H=\sum_{i= 1}^{A} t_{i} - T_{\rm{c.m.}}+\sum_{i> j}^{A} v_{ij}.
\end{equation}
 where the two-body interaction $(v_{ij})$ includes the central, spin-orbit, and Coulomb parts.
As the $NN$ interaction, for the central part, we use the Volkov No.2 effective potential \cite{VOL65}:
\begin{eqnarray}
V( r)=&&( W-MP^{\sigma}P^{\tau}+ BP^{\sigma}-HP^{\tau})\nonumber \\&&
\times
( V_{1}\exp(-r^{2}/ c_{1}^{2})+ V_{2}\exp(-r^{2}/ c_{2}^{2})),
\end{eqnarray}
 where $c_{1}=$ 1.01 fm, $c_{2}= 1. 8$ fm, $V_{1}=$ 61.14 MeV, $V_{2}=-60. 65$ MeV, $W= 1-M$ and $M= 0. 60$.
The singlet-even channel of the original Volkov interaction without the Bartlet ($B$) and Heisenberg ($H$) parameters has been known to be too strong, thus $B= H= 0.08$ is introduced to remove the bound state of two neutrons.
For the spin-orbit term, we introduce the G3RS potential \cite{G3RS-1,G3RS-2}:
\begin{equation}
V_{ls}= V_{0}( e^{-d_{1}r^{2}}-e^{-d_{2}r^{2}}) P(^{3}O) \bm{L} \cdot {\bm{S}},
\end{equation}
 where $d_{1}= 5. 0$ fm$^{-2},\ d_{2}= 2. 778$ fm$^{-2}$,
 $V_{0}= 2000$ MeV,  and $P(^{3}O)$ is a projection operator onto a triplet odd state.
The operator $\bm{L}$ stands for the relative angular momentum and $\bm{S}$ is the spin ($\bm{S_{1}}+\bm{S_{2}}$).
All of the parameters of this interaction were determined from the $\alpha+ n$ and $\alpha+\alpha$ scattering phase shifts \cite{OKA79}.

For the MCC calculation, we prepare the diagonal and transition densities.
The diagonal and transition densities are defined as;
\begin{eqnarray}
&&\rho_{Im,I'm'} (\bm{r}) \nonumber \\
&&= \Braket{\Phi^{J^{\pi} M} | \sum_{i=1} \delta (\bm{r}_{i}-\bm{R}_{\rm{c.m.}}-\bm{r}) | \Phi^{(J^{\pi})' M'} } \\
&&= \sqrt{4 \pi} \sum_{\lambda, \mu} (I' m' \lambda \mu | I m) \rho^{(\lambda)}_{I I'} (r) \mathscr{Y}^{*}_{\lambda \mu} (\hat{\bm{r}}), \label{eq:dens}
\end{eqnarray}
where $\mathscr{Y}_{LM}(\hat{\bm{r}}) = i^{L}Y_{LM}(\hat{\bm{r}})$.
Here, $(I' m' \lambda \mu | I m)$ denotes the Clebsch-Gordan coefficient.
$\bm{R}_{\rm{c.m.}}$ is the barycentric coordinate, and $m$ and $m'$ are the $z$-components of $I$ and $I'$, respectively.
The proton and neutron parts of the densities are separately obtained.
The wave function of the microscopic cluster model, $\Phi^{J^{\pi} M}$, is described as a linear combination of basis states as in Eq.~(\ref{mcmwf}), where coefficients for its linear combination are obtained by diagonalizing the Hamiltonian matrix.

\subsection{Microscopic coupled channel model}
Next, we explain the nuclear reaction calculation part.
We apply the calculated transition densities of the $^{9}$Li nucleus to the MCC calculations with the complex $G$-matrix interaction CEG07.
The coupled-channel (CC) equations for the radial component of the wave functions between colliding two nuclei, 
$\chi^{(J')}_{\alpha L}(R)$, for a given total angular momentum of the projectile-target scattering system $J'$ 
are written as,
\begin{equation}
\left[ T_{R}- E_{\alpha} \right] \chi^{(J')}_{\alpha L}(R) 
=-\sum_{\alpha', L'}{F^{(J')}_{\alpha L, \alpha' L'}(R) \chi^{(J')}_{\alpha' L'}(R)},
\label{eq:cc}
\end{equation} 
where $T_{R}$ denotes the kinetic-energy operator.
The suffix $\alpha$ denotes the channel number designated by the intrinsic spins of colliding two nuclei $I_1$ and $I_2$, the channel spin $S$ defined by the vector coupling of $I_1$ and $I_2$, and the sum of the excitation energies of the two nuclei $\epsilon_\alpha= \epsilon_1 + \epsilon_2$.
Namely, $\chi^{(J')}_{\alpha L}(R)$ is $\chi^{(J')}_{\alpha S(I_1 I_2) L}(R)$ if we write the indexes explicitly.
Here, we assign $\alpha = 0$ to the entrance (elastic) channel. 
$E_{\alpha} = E_{\rm{c.m.}}-\epsilon_\alpha$ is the center-of-mass (c.m.) energy of the projectile-target relative motion in the channel $\alpha$, where $E_{\rm{c.m.}}$ is the c.m. energy in the elastic channel.
The value $L$ is the orbital angular momentum for the relative motion between the two nuclei, which takes the values of $|J'-S|\leq L \leq J'+S$ for given $S$ and $J'$.
Thus, the scattering channel is defined by a set of $\alpha$ and $L$ for a given $J'$.
$F^{(J')}_{\alpha L, \alpha' L'}(R)$ represents the diagonal ($\alpha=\alpha'$ and $L=L'$) or coupling ($\alpha \neq \alpha'$ and/or $L \neq L'$) potential 
that is defined more explicitly~\cite{SYKT88, KAT02} as; 
\begin{eqnarray}
&& F^{(J')}_{\alpha L, \alpha' L'}(R) \equiv F^{(J')}_{\alpha S(I_1 I_2) L,\alpha' S'(I'_1 I'_2) L'}(R)  \nonumber \\
&=& \sum_{\lambda} i^{L + L' - \lambda} (-1)^{S + L' - J' - \lambda} \hat{L} \hat{L'}\,
W(S L S' L': J' \lambda) \nonumber \\
&\times& ( L 0 L' 0 | \lambda 0 ) U^{(\lambda)}_{\alpha S(I_1 I_2), \alpha' S'(I'_1 I'_2)}(R) ,
\label{eq:coupling}
\end{eqnarray}
where $\hat{L}=(2L+1)^{\frac{1}{2}}$, and $W(S L S' L: J' \lambda)$ denotes the Racah coefficient.

In Eq.~(\ref{eq:coupling}), $U^{(\lambda)}_{\alpha S(I_1 I_2), \alpha' S'(I'_1 I'_2)}(R)$ is the intrinsic component of the diagonal or coupling potential with the multipolarity of rank $\lambda$, that only contains nuclear structure information in channels $\alpha$ and $\alpha'$ and is irrelevant to the angular momenta $L$ and $J'$ associated with the projectile-target relative motion.
It consists of the Coulomb and nuclear parts,
\begin{eqnarray}
&&U^{(\lambda)}_{\alpha S(I_1 I_2), \alpha' S'(I'_1 I'_2)}(R) = \nonumber \\
&&V^{(\lambda,  \rm{Coul.})}_{\alpha S(I_1 I_2), \alpha' S'(I'_1 I'_2)}(R) + U^{(\lambda,  \rm{Nucl.})}_{\alpha S(I_1 I_2), \alpha' S'(I'_1 I'_2)}(R),
\end{eqnarray}
and they are obtained by the double folding of the Coulomb and nuclear parts of the $NN$ interaction, respectively, such as;
\begin{widetext}
\begin{eqnarray}
V^{(\lambda, \rm{Coul.})}_{\alpha S(I_1 I_2), \alpha' S'(I'_1 I'_2)}(R) &=&
\sqrt{4 \pi} \hat{S} \hat{S'} \hat{I_1} \hat{I_2}
\sum_{\lambda_1 \lambda_2},
\begin{Bmatrix}
I_1&I_2&S \\
I'_1&I'_2&S' \\
\lambda_1 &\lambda_2 &\lambda
\end{Bmatrix} \nonumber \\
&\times& \int{ \rho^{(\lambda_1, p)}_{I_1 I'_1}(r_1) 
               \rho^{(\lambda_2, p)}_{I_2 I'_2}(r_2)
               v^{(\rm{Coul.})}_{NN}(s) } 
\Big[ [       \mathscr{Y}_{\lambda_1}(\bm{\hat{r}}_1) 
       \otimes \mathscr{Y}_{\lambda_2}(\bm{\hat{r}}_2)]_{\lambda} 
       \otimes \mathscr{Y}_{\lambda}(\bm{\hat{R}}) \Big]_{00}
               d\bm{\hat{R}}d\bm{r}_1\bm{r}_2, 
\label{eq:coulfold} 
\end{eqnarray}
\begin{eqnarray}
U^{(\lambda, \rm{Nucl.})}_{\alpha S(I_1 I_2), \alpha' S'(I'_1 I'_2)}(R) &=&
\sqrt{4 \pi} \hat{S} \hat{S'} \hat{I_1} \hat{I_2}
\sum_{\lambda_1 \lambda_2}
\begin{Bmatrix}
I_1 &I_2 &S \\
I'_1&I'_2&S' \\
\lambda_1 &\lambda_2 &\lambda
\end{Bmatrix} \nonumber \\
&&\times \bigg \{ \int{ \rho^{(\lambda_1)}_{I_1 I'_1}(r_1) 
                  \rho^{(\lambda_2)}_{I_2 I'_2}(r_2)
                  v^{(\rm{D})}_{NN}(s, \rho, \epsilon ) } 
\Big[ [\mathscr{Y}_{\lambda_1}(\bm{\hat{r}}_1) \otimes \mathscr{Y}_{\lambda_2}(\bm{\hat{r}}_2)]_{\lambda} \otimes \mathscr{Y}_{\lambda}(\bm{\hat{R}}) \Big]_{00} d\bm{\hat{R}} d\bm{r}_1 d\bm{r}_2 \nonumber \\
&&\ \ \ \ \  + \int{ \hat{j}_1(k^{\rm{eff}}_F(p)s) \rho^{(\lambda_1 )}_{I_1 I'_1}(p)
           \hat{j}_1(k^{\rm{eff}}_F(t)s) \rho^{(\lambda_2 )}_{I_2 I'_2}(t) 
           v^{(\rm{EX})}_{NN}(s, \rho, \epsilon ) 
           } \nonumber \\
&&\ \ \ \ \ \ \ \times \exp{\{ \frac{i\bm{k}(R) \cdot \bm{s}}{\mu} \}}
               \Big[ [\mathscr{Y}_{\lambda_1}(\bm{\hat{p}}) 
               \otimes \mathscr{Y}_{\lambda_2}(\bm{\hat{t}})]_{\lambda} 
               \otimes \mathscr{Y}_{\lambda}(\bm{\hat{R}}) \Big]_{00} d\bm{\hat{R}} d\bm{p} d\bm{s}
\ \ \bigg \},
\label{eq:nuclfold}
\end{eqnarray}
\end{widetext}
where $\bm{s} = \bm{R} - \bm{r_1} + \bm {r_2}$, $\bm{p} = \bm{r}_1 + \frac{1}{2}\bm{s}$, and $\bm{t} = \bm{r}_2 - \frac{1}{2}\bm{s}$.
In this expression, the Wigner 9-$j$ symbol is introduced, and $E/A$ is the incident energy per nucleon. 
Here, $\mu = \frac{A_{1} A_{2}}{A_{1}+A_{2}}$, and $A_1$ and $A_2$ are the mass numbers of the projectile and target nuclei, respectively.
Note that $\rho^{(\lambda, p)}_{I I'}(r)$ with the superscript ($p$) in Eq.~(\ref{eq:coulfold}) represents the proton part of the density, which is used in the Coulomb part of the folding potential.
On the other hand, $v^{(D)}_{NN}$ and $v^{(EX)}_{NN}$ are the direct and exchange parts of the nuclear interaction, respectively, 
for which we adopt the complex $G$-matrix interaction CEG07b~\cite{CEG07, FUR09} and they are written as 
\begin{equation}
v_{\rm{D, EX}}=\pm \frac{1}{16}v^{00}+\frac{3}{16}v^{01}+\frac{3}{16}v^{10} \pm \frac{9}{16}v^{11},
\label{eq:vst}
\end{equation}
in terms of the spin-isospin components $v^{ST}$ ($S$ = 0 or 1 and $T$ = 0 or 1) of the CEG07 interaction. 

In the exchange part of Eq.~(\ref{eq:nuclfold}), $k(R)$ is the local momentum of the nucleus-nucleus relative motion defined by 
\begin{equation}
k^2(R)=\frac{2\mu}{\hbar^2}[E_{\rm{c.m.}}-{\rm{Re}}U^{(0, \rm{Nucl.})}_{0, 0}(R)
-V^{(0, \rm{Coul.})}_{0, 0}(R)], 
\label{eq:kkk}
\end{equation}
and the exchange part of the diagonal and coupling potentials is calculated self-consistently on the basis of the local energy approximation through Eq.~(\ref{eq:kkk}).
In Eq.~(\ref{eq:nuclfold}), $\hat{j}_1(k^{\rm{eff}}_F(x)s) \equiv \frac{3}{k^{\rm{eff}}_F(x)s}j_1(k^{\rm{eff}}_F(x)s)$, where $k^{\rm{eff}}_{F}$ is the effective Fermi momentum \cite{CAM78} defined by
\begin{equation}
k^{\rm{eff}}_{F} 
=\Big( (3\pi^2 \rho )^{2/3}+\frac{5C_{\rm{s}}[\nabla\rho]^2}{3\rho^2}
+\frac{5\nabla ^2\rho}{36\rho} \Big)^{1/2}, \;\; 
\label{eq:kf}
\end{equation}
and we adopt $C_{\rm{s}} = 1/4$ following Ref.~\cite{KHO01}. 
The exponential function in Eq.~(\ref{eq:nuclfold}) is approximated by the spherical Bessel function of rank 0, $j_{0} (\frac{Mk(R)s}{\mu})$,  following the standard prescription~\cite{BRI77,ROO77,BRI78,CEG83,CEG07,MIN10}.

We employ the so-called frozen-density approximation (FDA)~\cite{FUR09} for evaluating the local density $\rho$ in Eq.~(\ref{eq:nuclfold}).
In the FDA, the density-dependent $NN$ interaction is assumed to feel the local density defined as the sum of the densities of the projectile and target nuclei; 
\begin{eqnarray}
\rho &=& \rho^{(\rm P)}+\rho^{(\rm T)}.
\end{eqnarray}
In calculating the potentials, we use the average of the nucleon densities in the initial and final states for each nucleus~\cite{ITO01,KAT02};
\begin{eqnarray}
\rho^{(\rm P)} &=& \frac12 \left\{ \rho^{(0)}_{I_1 I_1}+ \rho^{(0)}_{I'_1 I'_1} \right\} \; , \\
\rho^{(\rm T)} &=& \frac12 \left\{ \rho^{(0)}_{I_2 I_2}+ \rho^{(0)}_{I'_2 I'_2} \right\} \;.
\end{eqnarray}
The local densities are evaluated at the position of each nucleon for the direct part and at the middle point of the interacting nucleon pair for the exchange part following the preceding works~\cite{KHO00,KAT02}. 
The FDA has widely been used also in the standard double folding model (DFM) calculations~\cite{KHO94, KHO97, KAT02, SAT79, FUR09} and it was proved that the FDA was the most appropriate prescription for evaluating the local density in the DFM calculations with realistic complex $G$-matrix interactions~\cite{FUR09}. 

Although the spin-orbit interaction between the nucleon-nucleon system is taken into account in the structure calculation of the $^9$Li nucleus, 
the spin-orbit potential between the $^{9}$Li and $^{12}$C nuclei is ignored in the present reaction calculation, which is shown to be negligible for the elastic and inelastic cross sections in Refs.~\cite{SAK86-so, ZAH96}. 
In addition, the magnetic-multipole (M1 and M3) transitions are also ignored in this paper, whose contributions will be discussed in the near future.

\vspace{3mm}
\section{Results}
\subsection{Energies, r.m.s. radii, and transition strengths}\label{structure}
\begin{figure}[t]
\begin{center}
\includegraphics[width=6.5cm]{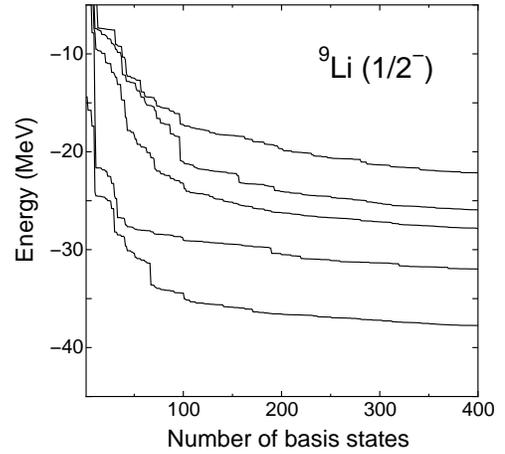}
\caption{\label{fig:01} Calculated total energy with the stochastic multi-configuration mixing method for the $\frac{1}{2}^{-}$ states.}
\end{center} 
\end{figure}
\begin{figure}[t]
\begin{center}
\includegraphics[width=6.5cm]{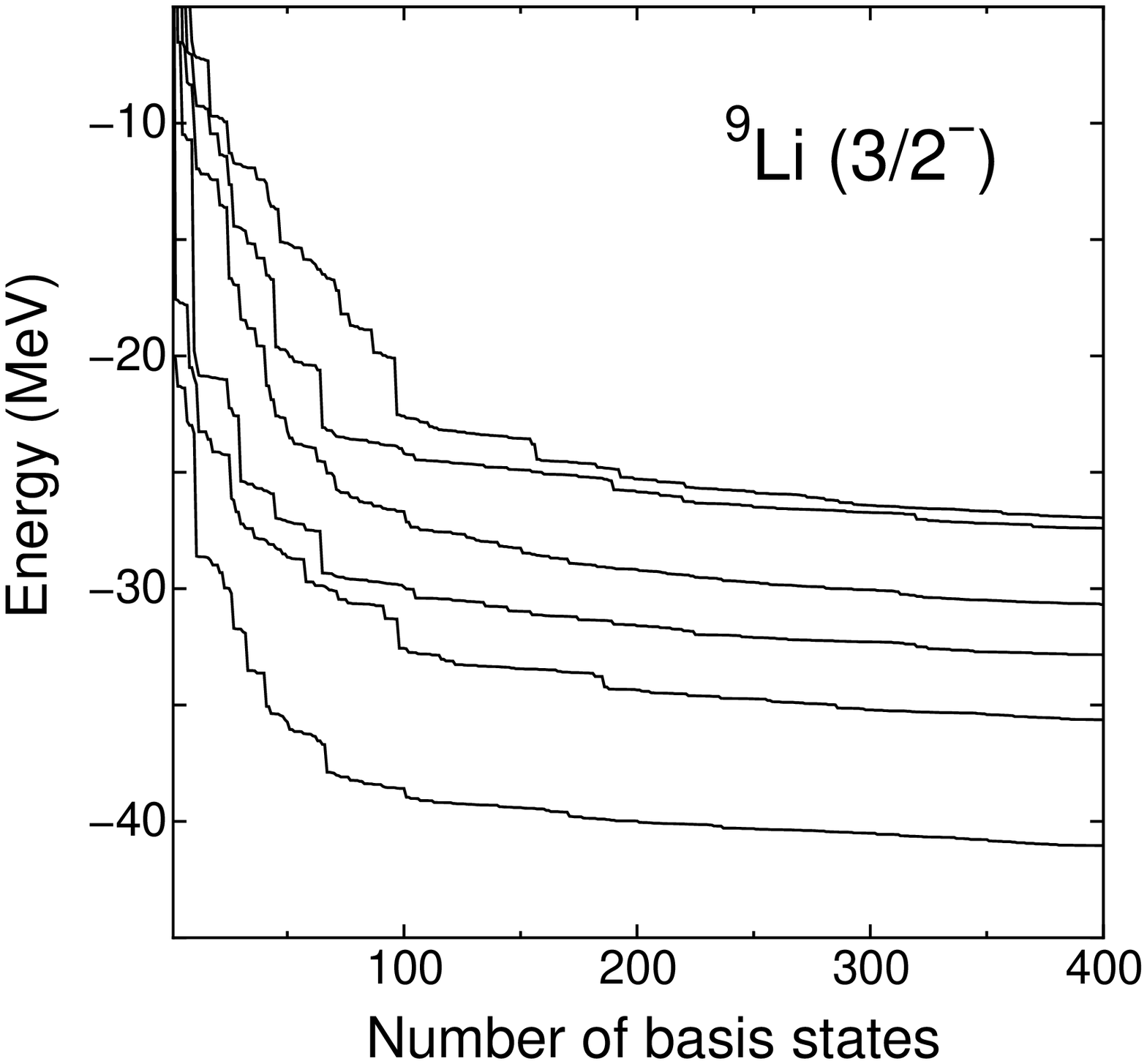}
\caption{\label{fig:02} Same as Fig.~\ref{fig:01} but for the $\frac{3}{2}^{-}$ states.}
\end{center} 
\end{figure}
\begin{figure}[t]
\begin{center}
\includegraphics[width=6.5cm]{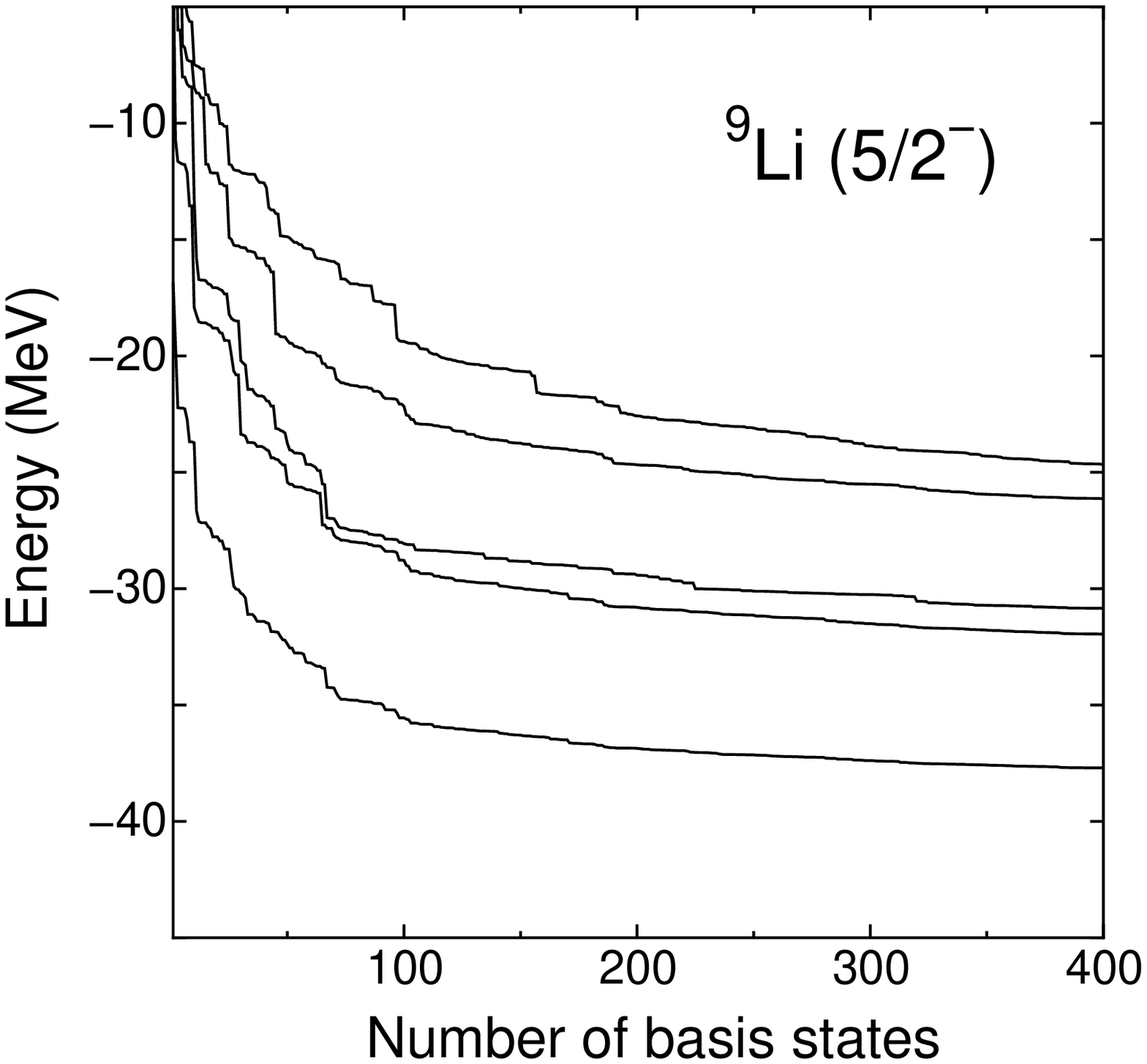}
\caption{\label{fig:03} Same as Fig.~\ref{fig:01} but for the $\frac{5}{2}^{-}$ states.}
\end{center} 
\end{figure}
\begin{figure}[t]
\begin{center}
\includegraphics[width=6.5cm]{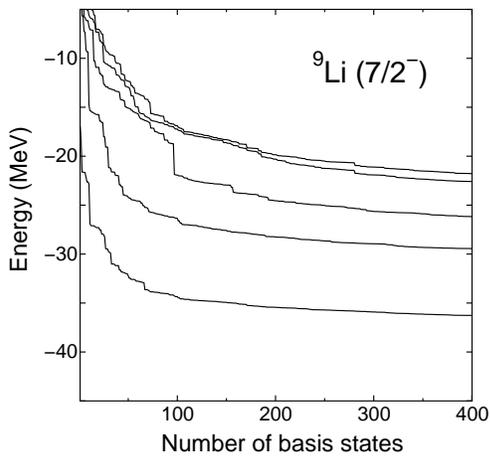}
\caption{\label{fig:04} Same as Fig.~\ref{fig:01} but for the $\frac{7}{2}^{-}$ states.}
\end{center} 
\end{figure}
We first calculate the total energies of the negative parity states for the $^{9}$Li nucleus using the wave function of the microscopic cluster model shown in Eq.~(\ref{mcmwf}).
The energy convergence of the ground and low-lying excited states is shown in Figs.~\ref{fig:01} to \ref{fig:04}. 
Here we take 400 basis states to confirm the energy convergence behavior. 
The calculated binding energy of the ground state, the $3/2^{-}_{1}$ state, is 41.04 MeV. 
This value is somewhat smaller than the experimental value, 45.34 MeV; however reasonable if we measure from the four-body threshold energy.
In experiments the neutron threshold opens at $E_x$ = 4.0639 MeV, and this value is almost the same in the present calculation. Above this threshold, in principle we have to impose the resonance condition for the obtained states when we distinguish resonance states and continuum states. 
However this is rather difficult and the present calculation adopts the bound state approximation. 
Nevertheless, the obtained states give standard radii and transition strengths; it can be considered that the obtained states are not continuum states but good candidates for the resonance states. 

\begin{figure}[t]
\begin{center}
\includegraphics[width=6.5cm]{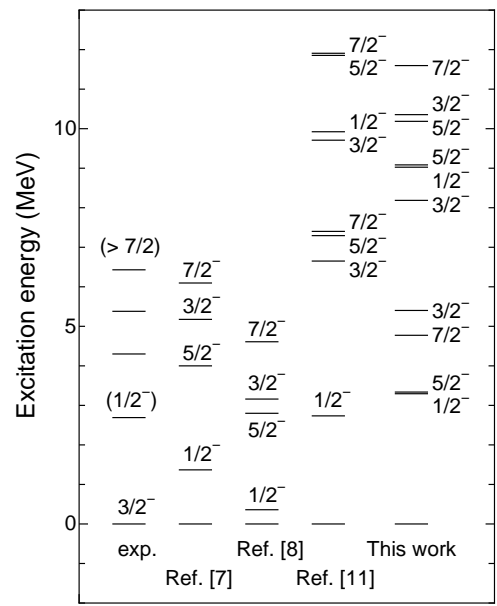}
\caption{\label{fig:05} Comparison of the calculated energies with experimental data and other calculation results. 
The experimental data is taken from Ref.~\cite{NNDC}.}
\end{center} 
\end{figure}
Next, we compare the calculated excitation energies with experimental data and other calculated results in Fig.~\ref{fig:05}.
All theoretical calculations give the ground and first excited states as 3/2$^{-}$ and 1/2$^{-}$, respectively; however in slightly higher excited region the predictions are different with each other.
In this work, the fist 5/2$^{-}$ state appears near the first 1/2$^{-}$ state, and the first 7/2$^{-}$ state is 
obtained under the second 3/2$^{-}$ state. 
For the experimental data, the spin assignment is not completed even for the first excitation state as in Ref.~\cite{NNDC}. 

\begin{table}[h]
\caption{Comparison of the calculated proton, neutron, and matter radii of the ground $3/2^-$ state of the $^{9}$Li nucleus together with the experimental data.}
\label{tab:01}
  \begin{tabular}{cccc} \hline
  & proton (fm) & neutron (fm) & matter (fm) \\ \hline  \hline
Exp.&  & &  \\ 
Ref.~\cite{TAN88} & 2.18(2) & 2.39(2) & 2.32(2) \\ 
Ref.~\cite{DOB06} & 2.11(4) & 2.59(9) & 2.44(6) \\ \hline 
Calc. &  &  &  \\ 
This work & 2.237 & 2.562 & 2.459 \\ 
Ref.~\cite{MYO12} & & & 2.46 \\ 
Ref.~\cite{ARA01} & 2.10 & 2.52 & 2.39 \\ 
Ref.~\cite{NAV98} & 1.946 &  &  \\ \hline
  \end{tabular}
\end{table}
Table~\ref{tab:01} shows the calculated root-mean-square (r.m.s.) radius of the ground $3/2^-$ state in this work compared with 
the other calculations and experimental data.
The present results give slightly large proton, neutron, matter radii in 
comparison with Ref.~\cite{TAN88}; however the difference is rather small.
On the other hand, the neutron and matter radii well agree with the experimental data of Ref.~\cite{DOB06}.

In the most recent report~\cite{FAL13}, the electric quadrupole transition strength is observed as 
$B$(E2; 1/2$^{-}_{1}$ $\rightarrow$ 3/2$^{-}_{1}$) = 6.8(3) $e^{2}$fm$^{4}$. 
In this work, the transition strength is obtained as $B$(E2) = 8.778 $e^{2}$fm$^{4}$. 
The theoretical value is slightly larger than the experimental one; however 
it can be considered that the obtained value reproduces the data fairly well.

\subsection{Elastic and inelastic cross sections}
Next, we introduce the MCC method and calculate the scattering cross sections for the $^{9}$Li + $^{12}$C system using the wave functions obtained above.
Here, we note that the imaginary part of the potential obtained by the folding is multiplied by a renormalization factor $N_W$ as
\begin{equation}
U = V + i N_{\rm{W}} W. 
\end{equation}
Here the $V$ and $W$ are the real and imaginary parts of the folding model potentials, respectively, and $N_{\rm{W}}$ is the only free parameter in the present CEG07 folding model.

In this paper, we fix this $N_{\rm{W}}$ value to reproduce the quasi-elastic scattering data for the $^{9}$Li + $^{12}$C system at $E/A =$ 60 MeV. 
In the quasi-elastic cross section, the excited states of the target $^{12}$C nucleus,
$2_{1}^{+}$ (4.44 MeV), $0_{2}^{+}$ (7.65 MeV), and $3_{1}^{-}$ (9.64 MeV) states,
are taken into account,  and the diagonal and transition densities of the $^{12}$C nucleus are taken 
from the 3$\alpha$-RGM (Resonance Group Method) calculation result~\cite{KAM81}. 
\begin{table}[h]
\caption{Total energies and states for the $^{9}$Li nucleus applied to the MCC calculation.}
\label{tab:02}
  \begin{tabular}{c|cccc} \hline
 & 1/2$^-$ & 3/2$^-$ & 5/2$^-$ & 7/2$^-$ \\ \hline  \hline
 4 & &-30.68& & \\ 
 3 &-27.84&-32.85&-30.85&-26.18 \\ 
 2 &-32.01&-35.64&-31.95&-29.44 \\ 
 1 &-37.74&-41.04&-37.70&-36.27 \\ \hline
  \end{tabular}
\end{table}
In addition, we take into account the excitation of the $^{9}$Li nucleus. 
Table~\ref{tab:02} shows the energies of the low-lying negative parity states of the $^{9}$Li nucleus, which are included in the MCC calculation. 
This is called  the ``full-CC calculation" in this paper.

\begin{figure}[t]
\begin{center}
\includegraphics[width=6.5cm]{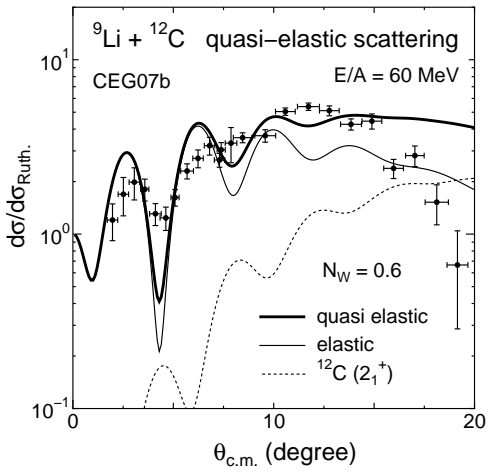}
\caption{\label{fig:06} Comparison of the experimental data~\cite{ZAH96} with the calculated quasi-elastic cross section with the $N_W$ value of 0.6.
The bold curves means the calculated quasi-elastic cross section.
The thin solid and dotted curves are the results of the elastic and inelastic cross sections, respectively.}
\end{center} 
\end{figure}
Figure~\ref{fig:06} shows the quasi-elastic scattering cross section for the $^{9}$Li + $^{12}$C system at $E/A =$ 60 MeV. 
The bold curve shows the calculated quasi-elastic cross section obtained by the incoherent sum of the elastic and inelastic cross sections.
The solid and dotted curves show the calculated elastic and inelastic scattering cross sections, respectively.
The calculated quasi-elastic cross section with $N_W =$ 0.6 reproduce the data, except for the most backward angles.
Then, we fix to $N_W =$ 0.6 and discuss the $^{9}$Li structure in the inelastic scattering angular distribution only up to 15 degrees.

\begin{figure}[t]
\begin{center}
\includegraphics[width=6.5cm]{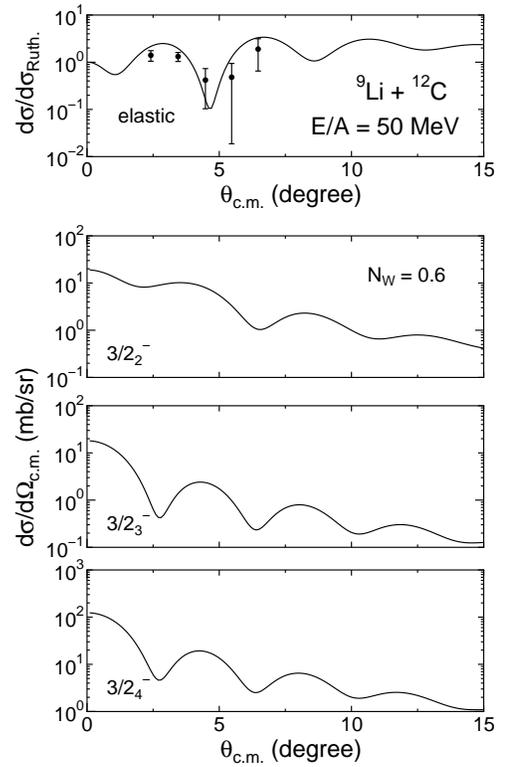}
\caption{\label{fig:07} Calculated elastic and inelastic (to $3/2_2^-$, $3/2_3^-$, and $3/2_4^-$ states) cross sections for the $^{9}$Li + $^{12}$ C at $E/A =$ 50 MeV.
The experimental data is taken from Ref.~\cite{PET03}.}
\end{center} 
\end{figure}
Next, we calculate the elastic and inelastic cross sections for the $^{9}$Li + $^{12}$C system at $E/A =$ 50 MeV.
The calculated elastic cross sections well reproduce the data as shown in Fig.~\ref{fig:07}.
In the transition from a 3/2$^{-}$ state to another 3/2$^{-}$ state, the quadrupole and monopole transitions compete with each other.
Here, it can be seen that the angular distribution of the inelastic cross section to the 3/2$^{-}_{2}$ final state is clearly different from those of the other 3/2$^{-}$ states; the angular distribution of the 3/2$_2^-$ final state clearly shows the pattern of quadrupole transition, and those of the 3/2$_3^-$ and 3/2$_4^-$ final states show the patterns of monopole transition.

\begin{figure}[t]
\begin{center}
\includegraphics[width=6.5cm]{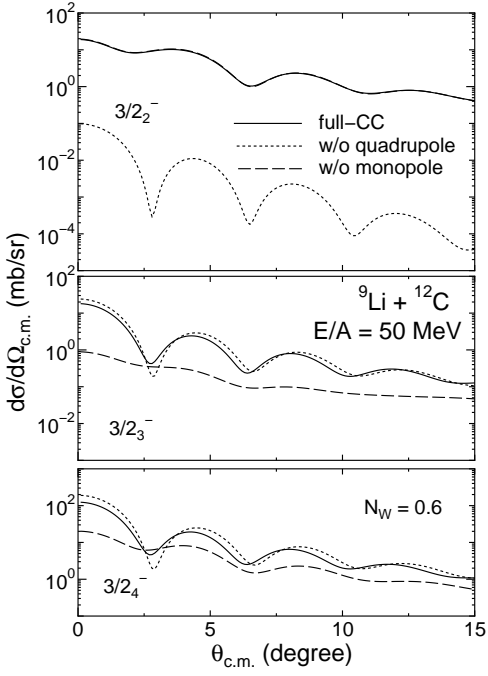}
\caption{\label{fig:08} Decomposition of the calculated inelastic cross sections into monopole and quadrupole components.
The solid curves are the same to Fig.~\ref{fig:07}.
The dotted and dashed curves are the results without the quadrupole and monopole transitions, respectively.}
\end{center} 
\end{figure}
In order to confirm the situations, we decompose the calculated inelastic cross sections into the monopole and quadrupole components, which are shown in Fig.~\ref{fig:08}.
The solid curves are the calculated inelastic cross sections, and the dotted and dashed curves are the calculated inelastic cross sections without the quadrupole and monopole transitions, respectively.
For the 3/2$^{-}_{2}$ state, the quadrupole component plays a dominant role in the calculated inelastic cross section. 
Here, we note that the dashed curve well agrees with the solid curve and it is difficult to distinguish the curves in the upper panel.
On the other hand, the calculated inelastic cross section is mainly determined by the monopole component in the results of 
other 3/2$^{-}$ final states. 
When the monopole transition plays a dominant role in the inelastic cross section, the quadrupole transition strength also becomes large passably.
Here the large monopole transition implies the increase of the size of the nucleus, namely, the development of the $\alpha$ + $t$ cluster structure~\cite{YAM08}. 
The large quadrupole transition strength can be interpreted as the result of this development of this cluster structure.

\begin{figure}[t]
\begin{center}
\includegraphics[width=6.5cm]{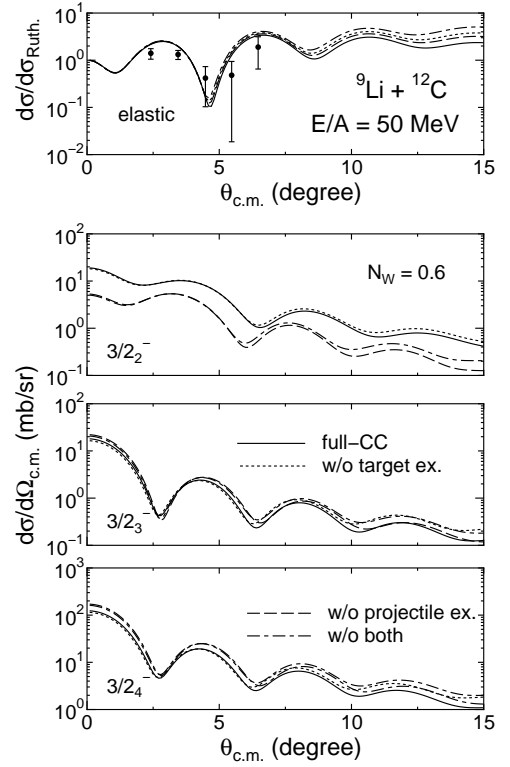}
\caption{\label{fig:09} Mutual-excitation and multi-step effects on the elastic and inelastic cross sections.
The solid curves are the same as Fig.~\ref{fig:07}.
The dotted, dashed, and dot-dashed curves are the results without the target, projectile, both excitation effects, respectively.}
\end{center} 
\end{figure}
In addition, we investigate the mutual-excitation and multi-step effects caused by the projectile and target excitations.
In Fig.~\ref{fig:09}, the solid curves show the full-CC calculation results, and the dotted and dashed curves are the results without the target and projectile excitation effects, respectively.
The dot-dashed curve without both excitations includes no channel coupling effect for the elastic cross section.
For the inelastic cross section, the dot-dashed curves is obtained by the two-channel calculation.
The calculated results suggest that the target excitation has a minor role for the elastic and inelastic scatterings of 
the $^{9}$Li + $^{12}$C system at $E/A =$ 50 MeV in this angular distribution.
Here, we can see a drastic change after considering the couplings in the inelastic scattering cross section to the 3/2$_2^-$ final state; the change of the absolute value can be explained only by taking into account the multi-step reaction effect corresponding to the projectile excitation ($^{9}$Li$^*$).
We discuss this exotic 3/2$_2^-$ state in detail in the next subsection.

\subsection{Discussion for the 3/2$_2^{-}$ state}
\begin{figure}[t]
\begin{center}
\includegraphics[width=6.5cm]{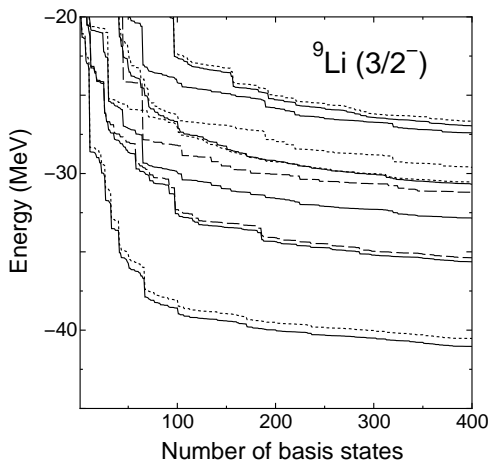}
\caption{\label{fig:10} Decomposition of the calculated 3/2$^-$ states into the $K =$ 1/2 and 3/2 components.
The solid curves are the same as Fig.~\ref{fig:02}.
The dotted and dashed curves are the results only with $K$ = 1/2 and $K$ = 3/2, respectively}
\end{center} 
\end{figure}
In order to investigate the 3/2$^{-}_{2}$ state in detail, we first return back to the nuclear structure calculation introduced in subsection~\ref{structure}. 
Here we compare the calculated binding energies by fixing each $K$ quantum number. 
In principle the $K$ quantum number cannot be determined uniquely for non-axial symmetric nuclei as discussed in Ref.~\cite{BEN08}; however fixing the $K$ quantum numbers and comparing the results is a good prescription to investigate the character of the state from the theoretical side.
Figure~\ref{fig:10} shows the calculated 3/2$^{-}$ states of the $^{9}$Li nucleus.
The solid curves are the same as Fig.~\ref{fig:02}, the full $K$-mixing calculation, and the dotted and dashed curves are the results obtained by fixing to $K = 1/2$ and $K = 3/2$, respectively.
The ground state is well described with the result of $K =$ 1/2, whereas the 3/2$_2^-$ state is with $K = 3/2$.

\begin{table}[h]
\caption{$Q_J$ moments [($e$)fm$^2$] of the ground and 3/2$_2^-$ states for the proton and neutron parts.}
\label{tab:03}
  \begin{tabular}{ccc} \hline
               & proton ($e$fm$^2$)& neutron (fm$^2$) \\ \hline  \hline
g.s.           & -2.651 & -3.333  \\ 
3/2$_2^-$ & 3.221 & 4.343 \\  \hline 
  \end{tabular}
\end{table}
The $Q_J$ moments for the ground and second 3/2$^{-}$ states are calculated and shown in Table~\ref{tab:03}. 
The obtained value for the proton part of the ground state well reproduces the experimental value, -2.53(9) $e$fm$^2$~\cite{ARN88}.
In addition, the value is in good agreement with the other theoretical calculations, -2.74 $e$fm$^2$~\cite{ARA01} and -2.7 $e$fm$^2$~\cite{ENY01}. 
If we consider the ideal case and each state has a good $K$ quantum number ($K = 1/2$ for the ground $3/2_1^-$ state and $K = 3/2$ for the $3/2_2^-$ state), the intrinsic quadrupole moment, $Q_0$, can be obtained from the $Q_J$ moment as follows;
\begin{equation}
Q_J = \frac{3K^2 - J(J+1)}{2J(J+1)}Q_0.
\end{equation}
Using this equation, both $Q_0$ values for the ground and 3/2$_2^-$ states are obtained to be positive. 
It means that the proton and neutron in the ground and 3/2$_2^-$ states prefer a prolate like deformation.

\begin{table}[h]
\caption{Expectation values of $\braket{\bm{L} \cdot \bm{S}}$ and $\braket{\bm{S}^2}$ for the neutron part of the $^{9}$Li and $^{10}$Be nuclei.}
\label{tab:04}
  \begin{tabular}{ccc} \hline
               & $\braket{\bm{L} \cdot \bm{S}}$ & $\braket{\bm{S}^2}$ \\ \hline \hline
$^{9}$Li&& \\
g.s.              & 1.064 & 0.5076 \\ 
3/2$_{2}^{-}$ & 0.08630 & 0.08054 \\ \hline \hline
$^{10}$Be&& \\
2$_{1}^{+}$     & 0.8901 & 0.5386 \\
2$_{2}^{+}$     & 0.1303 & 0.1056 \\ \hline
  \end{tabular}
\end{table}
Next we discuss the configurations of two valence neutrons rotating around the $\alpha + t$ core.
We calculate the expectation values of $\bm{L} \cdot \bm{S}$ and $\bm{S}^2$ operators for the neutron part.
Here $\bm{L} \cdot \bm{S}$ is a sum of one-body spin-orbit operators, $\sum_i \bm{l}_i \cdot \bm{s}_i$ (here $\bm{l}_i$ and $\bm{s}_i$ are orbital angular momentum and spin operators for the $i$-th neutron, respectively), and $\bm{S}^2$ is a sum of two-body spin operators, $\sum_{i,j} \bm{s}_i \cdot \bm{s}_j$, where summations are for the neutrons in both cases.
The contribution from the neutrons in the $\alpha + t$ core is zero due to the symmetry of the clusters, and we can discuss the contribution only from the two valence neutrons.
For instance, when two neutrons occupy the $p_{3/2}$ and $p_{1/2}$ orbits, the value of $\Braket{\bm{L} \cdot \bm{S}}$  is $-0.5$, which is simply the sum of the eigenvalues of $\bm{l} \cdot \bm{s}$ for those orbits, 0.5 and $-1.0$, respectively.
The calculated values are shown in Table~\ref{tab:04}. 
For the ground state, the $\Braket{\bm{L} \cdot \bm{S}}$ value is almost 1, and two valence neutrons are considered to dominantly occupy the $p_{3/2}$ orbits.
For the second 3/2$^{-}$ state, the $\Braket{\bm{L} \cdot \bm{S}}$ value is almost 0, which suggests the mixing of di-neutron configuration, since di-neutron is a spin-zero state and has an eigenvalue of zero for the one-body spin-orbit operator $\bm{L} \cdot \bm{S}$. 
We can confirm this situation from the calculated $\Braket{\bm{S}^2}$ value, which is 0.5 for the ground state but almost zero for the second 3/2$^{-}$ state. 
In the $3/2_2^-$ state, two neutrons are considered to have spin-zero, which also implies the mixing of di-neutron components in this state.
The excitation energy of the $3/2_2^-$ state is about 6 MeV, which is near to the two-neutron threshold energy of $^{9}$Li.
The energy position of the $3/2_2^-$ state also supports the mixing of the di-neutron components.

Here, we briefly mention about the similarity to the $^{10}$Be case. 
The $^{10}$Be nucleus is well described by an $\alpha$ + $\alpha$ + $n$ + $n$ model, and there appear two rotational bands originating from the $K = 0$ and $K = 2$ configurations of the two valence neutrons around the $\alpha$ + $\alpha$ core~\cite{ITA00a}.
The first $2^+$ state belongs to the ground band and dominantly has $K = 0$, whereas the second $2^+$ state belongs to the side band of the ground band and has dominantly $K = 2$.
In the second $2^+$ state, di-neutron configuration mixes, since the state is close to the two-neutron threshold energy~\cite{ITA00b}.
As a result, $K$ mixing effect is large for the second $2^+$ state, which means a mixing of the triaxial ($\alpha$ + $\alpha$ + di-neutron) components.
The similarity between the $^9$Li and $^{10}$Be nuclei was discussed in Ref.~\cite{SUH10} by comparing the energy surfaces and structures on the $\beta$-$\gamma$ plane of both nuclei.
They found that the cluster features in the $^{9}$Li nucleus are analogous to those in the $^{10}$Be nucleus by replacing one $\alpha$ cluster in the $^{10}$Be nucleus to a $t$ cluster in the $^{9}$Li nucleus.
Here we also find the similarity; in Table~\ref{tab:04}, the $\Braket{\bm{L} \cdot \bm{S}}$ and $\Braket{\bm{S}^2}$ values are almost the same for $^9$Li($3/2_{1}^{-}$) and $^{10}$Be($2^{+}_{1}$) ($\Braket{\bm{L} \cdot \bm{S}}$ and $\Braket{\bm{S}^2}$ values are 1.0 and 0.5 for both states, respectively).
The values are almost similar to the ground $0^+$ state of the $^{10}$Be nucleus, thus the two neutron configurations are almost the same in the ground states of both nuclei.
In addition, both of the $^{10}$Be($2_2^+$) and $^9$Li($3/2_2^-$) states have very small expectation values of $\Braket{\bm{L} \cdot \bm{S}}$ and $\Braket{\bm{S}^2}$, and mixing of di-neutron components is considered to be important. 

\begin{figure}[t]
\begin{center}
\includegraphics[width=6.5cm]{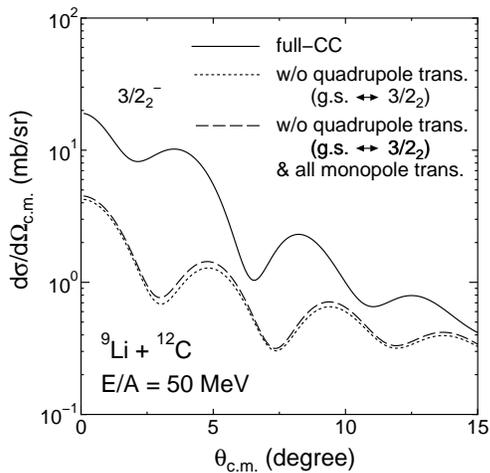}
\caption{\label{fig:11} Calculated inelastic cross section of the 3/2$_2^-$ state.
The curves are explained in the text.}
\end{center} 
\end{figure}
\begin{table}[h]
\caption{Monopole and quadrupole transition strengths (($e^2$)fm$^4$) between the low-lying states for the $^{9}$Li nucleus.}
\label{tab:05}
  \begin{tabular}{cccc} \hline
               & $B$(E$\lambda$) ($e^2$ fm$^4$)& neutron (fm$^4$)& $B$(IS$\lambda$) (fm$^4$) \\ \hline \hline
$\lambda = 2$&&& \\
g.s. $\rightarrow$ 1/2$_{1}^{-}$ & 4.389 & 5.258 & 19.26 \\ 
1/2$_{1}^{-}$ $\rightarrow$ 3/2$_{2}^{-}$ & 0.1583 & 14.52 & 17.71 \\ 
g.s. $\rightarrow$ 3/2$_{2}^{-}$ & 0.3490 & 2.130 & 0.7547 \\ 
g.s. $\rightarrow$ 5/2$_{1}^{-}$ & 0.08766 & 16.68 & 19.19 \\
5/2$_{1}^{-}$ $\rightarrow$ 3/2$_{2}^{-}$ & 2.566 & 3.210 & 11.52 \\  \hline \hline
$\lambda = 0$&&& \\
g.s. $\rightarrow$ 3/2$_{2}^{-}$ & 0.0004118 & 0.001180 & 0.0001976 \\  \hline 
  \end{tabular}
\end{table}
Finally, we investigate the angular distribution of the inelastic scattering to the 3/2$_2^-$ state.
The calculated result is shown in Fig.~\ref{fig:11}, where the solid curve shows the full-CC calculation result and the dotted curve shows the result without the quadrupole transition between the ground state and the 3/2$_2^-$ state.
The dashed curve shows the result removing all the monopole transition from the dotted curve.
The inelastic cross section of the 3/2$_2^-$ state is determined not only by the quadrupole transition between the ground state and the 3/2$_2^-$ state but also by the quadrupole transitions between the 3/2$_2^-$ state and other states as shown in Figs.~\ref{fig:09} and~\ref{fig:11}.
In particular, the quadrupole transitions through the 1/2$_1^-$ and 5/2$_1^-$ states play a major role to increase the inelastic cross section.
The transition strengths are summarized in Table~\ref{tab:05}.
In addition, we note that the dashed curve looks similar to the pattern of the monopole angular distribution, although all the monopole transitions are removed.
This result implies that the two-step quadrupole transition has similar effect to the monopole transition in the angular distribution. 
However, it is difficult to confirm such situation directly, because direct transition reaction from the entrance channel is considered to have a dominant role for all states.

\section{Summary}
We have investigated low-lying negative parity states of the $^{9}$Li nucleus using a unified framework of microscopic structure and reaction models, where we have used the same $^{9}$Li wave function for the structure and reaction calculation parts.
The $^{9}$Li wave function is fully antisymmetrized and consists of $\alpha$ + $t$ + $n$ + $n$ four bodies, and the low-lying 1/2$^{-}$, 3/2$^{-}$, 5/2$^{-}$, and 7/2$^{-}$ states are obtained by the stochastic multi-configuration mixing method.
The configuration of each basis state is randomly generated, and the total wave function is described by the superposition of the Brink-type wave functions.
The eigenstates of the Hamiltonian were obtained by diagonalizing the Hamiltonian, and bound state approximation is used for the unbound states.
Using these wave functions, the elastic and inelastic cross sections were calculated in the framework of the microscopic coupled channel method with the complex $G$-matrix interaction CEG07.
The calculated inelastic cross sections to the excited 3/2$^{-}$ states show the different angular distribution patterns, arising from the competition between the monopole and quadrupole transitions in the excitation.
It is found that the quadrupole excitation, rather than the monopole excitation, is dominant for the 3/2$^{-}_{2}$ state, contrary to the cases of other low-lying $3/2^-$ states.
In addition, we investigated the mutual-excitation and multi-step effects on the elastic and inelastic cross sections. 
For the 3/2$^{-}_{2}$ state, the sizable multi-step effect has been seen, but the multi-step effect is not important for the other cross sections.

In order to investigate the properties of the 3/2$^{-}_{2}$ state in detail, we performed a structure calculation restricted to $K =$ 1/2 and $K =$ 3/2; the ground and 3/2$_2^-$ states are well descried by $K =$ 1/2 and $K =$ 3/2, respectively.
From the calculated $Q_0$ moments, the first and second $3/2^{-}$ states have 
been shown to prefer a prolate like deformation.
In addition, the expectation values of 
$\bm{L} \cdot \bm{S}$ and $\bm{S}^2$ were calculated and compared with those of the $^{10}$Be nucleus.
From these values, we can find that the valence neutrons have almost the same $(p_{3/2})^2$ configuration in the both ground states of 
the $^9$Li and $^{10}$Be nuclei.
For the excited states, di-neutron components are considered to be mixed in the second $3/2^-$ state of the $^{9}$Li nucleus
as well as the second $2^{+}$ state of the $^{10}$Be nucleus, which implies the contribution of the triaxial ($\alpha$ + $t$ + di-neutron) components. 
Finally, we have investigated the calculated inelastic cross section of second 3/2$^{-}$ state in detail.
Not only the transition from the ground state to the 3/2$_2^-$ state but also other quadrupole transitions contribute to the cross section.
Especially, the multi-step transitions through the 1/2$_{1}^{-}$ and 5/2$_{1}^{-}$ states have an important role to determine the second 3/2$^{-}$ cross section.

In the present paper, detail discussion has been made only for the second 3/2$^{-}$ state. 
However other low-lying excited states of the $^{9}$Li nucleus are simultaneously obtained, and
their properties will be investigated in the near future.
Furthermore, these excited states of the $^{9}$Li nucleus 
are expected to contribute to the reaction dynamics of the $^{11}$Li nucleus, which will be also investigated.
The study of the core nucleus ($^{9}$Li) itself, as in this paper, is crucial for such study on the coexistence nature of weak (valence neutrons) and strong (core nucleus) bindings of unstable nuclei.

\section{Acknowledgment}
The authors would like to thank to Professors K. Matsuyanagi and Y. Kanada-En'yo for useful advices. 
The authors would like to thank to Professors A.~M.~Moro and K.~Ogata for encouraging comments.


\end{document}